\RequirePackage{fixltx2e}
\documentclass[8.5pt,twoside,twocolumn]{article}
\usepackage{graphicx,subfigure,amsmath} 

\begin{document}

\twocolumn[
  \begin{@twocolumnfalse}
\noindent\LARGE{\textbf{Characterizing the greater-than-bulk melting behaviour of Ga-Al nanoalloys}}
\vspace{0.6cm}

\large{\textbf{Udbhav Ojha and Nicola Gaston}}\vspace{0.3cm}

\large{The MacDiarmid Institute for Advanced Materials and Nanotechnology, Victoria University of Wellington, P.O. Box 600, Wellington 6012, New Zealand.\newline
\centering{E-mail: nicola.gaston@vuw.ac.nz}}
\newline

\noindent \normalsize{Phase diagrams are the most reliable way of predicting phase changes in a system. In this work, we create the phase diagram of the Ga\textsubscript{(20-x)}Al\textsubscript{x}\textsuperscript{+} bimetallic cluster system, for which the monometallic clusters display greater-than-bulk melting behaviour. Employing first-principles Born-Oppenheimer molecular dynamics in the microcanonical ensemble, we describe the solid-liquid-like phase transition in two distinct compositions Ga\textsubscript{11}Al\textsubscript{9}\textsuperscript{+} and Ga\textsubscript{3}Al\textsubscript{17}\textsuperscript{+}. The clusters show peaks in specific heat at 824 K and 922 K respectively which is above the corresponding bulk alloy melting temperatures. Mean squared displacements, diffusion coefficients and atoms-in-molecules analysis emphasise the differences between the environments of the internal and surface atoms. An Arrhenius-like description of the `ease' with which internal and surface sites (or atoms) can exchange positions with temperature is used to quantify the \textit{greater-than-bulk melting character} which corroborates the view of a Ga\textsubscript{(20-x)}Al\textsubscript{x}\textsuperscript{+} cluster being a remnant of the corresponding bulk alloy material with additional characteristics specific to its size and composition.}
\vspace{0.5cm}
 \end{@twocolumnfalse}
  ]

\section{Introduction}
Nanoalloys comprising two or more metal atoms within a finite framework offer novel physical and chemical properties significantly different from their bulk alloy counterparts\cite{nanoalloys}. The sensitivity of their properties to system size and composition gives ample room for fine tuning as desired for various applications in catalysis, optics and electronics\cite{Johnstonbook}. Understanding the effects of size, structure, composition, chemical ordering and temperature on the emergent behaviour of nanoalloys thus plays a key role in their fabrication and functional optimization.   

Although a lot of scientific literature within current nanoscience addresses the role of ground state nanoalloy configurations on the observed properties\cite{nanoalloys,Johnstonbook, nanoalloy_book, FerForJohn}, effects at finite-temperature have received less attention\cite{calvo_nanoalloy,polak}. One such property is the solid-liquid-like phase transition temperature which has already been found, both experimentally\cite{jesser1,jesser2} and in theoretical simulations\cite{th1,th2,th3,th4}, to depend on the size and composition of nanoalloys. Moreover, the phase transition behaviour changes qualitatively as well; different melting stages (premelting) were observed in Cu-Au system\cite{lopez} in contrast to Ag-Au bimetallic clusters\cite{silvergold} which showed a single stage melting. Complex structural transformations, including switching among homotops before complete melting, have also been observed\cite{struc1,struc2}. 

Compositional dependence of the phase transition temperature in alloys is studied through the construction of phase diagrams. Extension of the phase diagram to nanoscale materials for predicting the solid-liquid-like phase transition temperature has been already attempted\cite{th1,pd2,pd3,pd4}. Models typically used for bulk phases\cite{bulk1,bulk2} are unsuitable for these systems and hence, sensitive first-principles treatment of the electronic structure is necessary. In our previous study\cite{self}, we showed that density functional theory (DFT) was able to capture the solid-liquid-like phase transition temperature of Ga\textsubscript{19}Al\textsuperscript{+} in excellent agreement with experiments\cite{jarroldGa19Al}. In this work, we explore two more cluster sizes in the Ga\textsubscript{(20-x)}Al\textsubscript{x}\textsuperscript{+} cluster series: Ga\textsubscript{11}Al\textsubscript{9}\textsuperscript{+} and Ga\textsubscript{3}Al\textsubscript{17}\textsuperscript{+}, and attempt to create a phase diagram of Ga-Al nanoalloy system using the phase transition temperature data of monometallic phases, Ga\textsubscript{20}\textsuperscript{+} and Al\textsubscript{20}\textsuperscript{+} respectively\cite{krista1,selfAl}. 

Cationic gallium clusters have been observed to show melting-like transition at temperatures significantly above the bulk melting temperature of gallium\cite{jarrold_secondorder, jarrold_PRB, jarrold_GaJACS, jarrold_GaPRL,Krista_PRB}. However, a recent study by Jarrold and co-workers\cite{Jarrold60} have found the  gallium cluster comprising 95 atoms to be the first cluster in the size range 60-183 atoms to melt below the bulk melting temperature. Although not observed in experiments due to limitations on the temperature range\cite{Al16-48}, greater-than-bulk melting behaviour has also been predicted for aluminium clusters\cite{selfAl}. An aluminium atom, being isoelectronic and having a similar ionization potential (IP), has a vacant \textit{d}-shell and is 5.4\% smaller in atomic radius in comparison to a Ga atom. Moreover, the overlap of valence \textit{s} and \textit{p} orbitals is different for both the metals since the \textit{4s} band of Ga is shifted to a higher binding energy in comparison to the \textit{3s} band of Al\cite{Eberhardt}. These differences raise the question as to how the thermodynamics of Ga-Al clusters change with composition in the nano-regime, and also how is it different from the corresponding bulk Ga-Al alloy system. The aim of this report is to answer the above question and also to understand further the differences between the behaviour of the internal and surface atoms of the cluster and how they affect the phase transition, as has been observed in Ga\textsubscript{20}\textsuperscript{+} and Al\textsubscript{20}\textsuperscript{+} clusters\cite{selfAl}. The organisation of the work is as follows: section 2 describes the computational method adopted for this study. In section 3 we present the results obtained using various statistical tools and indices to explain the spatial and temporal variations occurring during the phase transition of Ga$_{11}$Al$_{9}^+$ and Ga$_{3}$Al$_{17}^+$ clusters and go on to complete the phase diagram of the 20-atom Ga-Al system. Finally we summarise and conclude in section 4.

\section{Computational details}
We adopted a similar methodology for studying the melting transitions in Ga\textsubscript{11}Al\textsubscript{9}\textsuperscript{+} and Ga\textsubscript{3}Al\textsubscript{17}\textsuperscript{+} clusters as has been described in our previous works\cite{self, selfAl}. Density functional theory (DFT) calculations were performed using the plane waves basis set based code Vienna \textit{ab-initio} simulation package (VASP)\cite{vasp1,vasp2,vasp3,vasp4}. Large core (i.e. only \textit{3s$\rm^{2}$3p$\rm^{1}$} and \textit{4s$\rm^{2}$4p$\rm^{1}$} electrons were considered as valence for Al and Ga respectively) projector-augmented wave (PAW)\cite{paw1,paw2} pseudo-potentials in combination with the generalized gradient approximation in the Perdew-Wang form (GGA-PW91)\cite{PW91a,PW91b} were used. The starting structures for Ga\textsubscript{11}Al\textsubscript{9}\textsuperscript{+} and Ga\textsubscript{3}Al\textsubscript{17}\textsuperscript{+}  clusters (as shown in Fig.~\ref{structure}) were determined using the stacked plane (SP)\cite{krista1,self} Ga$_{20}$ structure. Each of the 20 available sites in the SP configuration of Ga$_{20}$ was substituted by a single Al atom and based on the calculated free energy of the cluster, the sites were numbered in ascending order. The first nine and seventeen sites served as the positions for Al atoms in Ga\textsubscript{11}Al\textsubscript{9}\textsuperscript{+} and Ga\textsubscript{3}Al\textsubscript{17}\textsuperscript{+} respectively.  

Born-Oppenheimer molecular dynamics (BOMD) as implemented in VASP in addition to parallel tempering (PT) was used to perform the molecular dynamics simulations in the microcanonical ensemble. 19 different temperatures between 250 K and 1150 K at 50 K intervals were chosen to capture the solid-liquid-like phase transition. Relative convergences of the canonical and microcanonical specific heat curves obtained using the multiple histogram (MH) method\cite{CalvoPTMD, krista1}, ascertained using the sliding window analysis\cite{krista1}, were achieved after a total run time of 160 ps (step size 2 fs) for Ga\textsubscript{11}Al\textsubscript{9}\textsuperscript{+} and 46.2 ps (step size 0.8 fs) for Ga\textsubscript{3}Al\textsubscript{17}\textsuperscript{+}. A smaller step size was used in Ga\textsubscript{3}Al\textsubscript{17}\textsuperscript{+} to counter the increase in energy during microcanonical runs. Parallel tempering (PT) helps to avoid the dependence of thermodynamics on the starting structure and the  possibility of the structure getting trapped in one of the local minimas in the potential energy surfaces of Ga\textsubscript{11}Al\textsubscript{9}\textsuperscript{+} and Ga\textsubscript{3}Al\textsubscript{17}\textsuperscript{+}. After every 100 time steps, each of the 19 different microcanonical runs was stopped. Two configurations were randomly selected and based on the calculated acceptance probability\cite{Calvo_PTMD_new}, swapping was either accepted or rejected. The spatial charge distribution analysis was performed using freely available software\cite{henkelman} based on the quantum theory of atoms-in-molecules (QTAIM)\cite{bader}.       


\section{Results and Discussion}

\subsection{Phase transition in Ga$_{11}$Al$_{9}^+$ and Ga$_{3}$Al$_{17}^+$}
\begin{figure}[t]
\centering
\includegraphics[trim = 0cm 0cm 0cm 0cm,clip=true,width=\columnwidth]{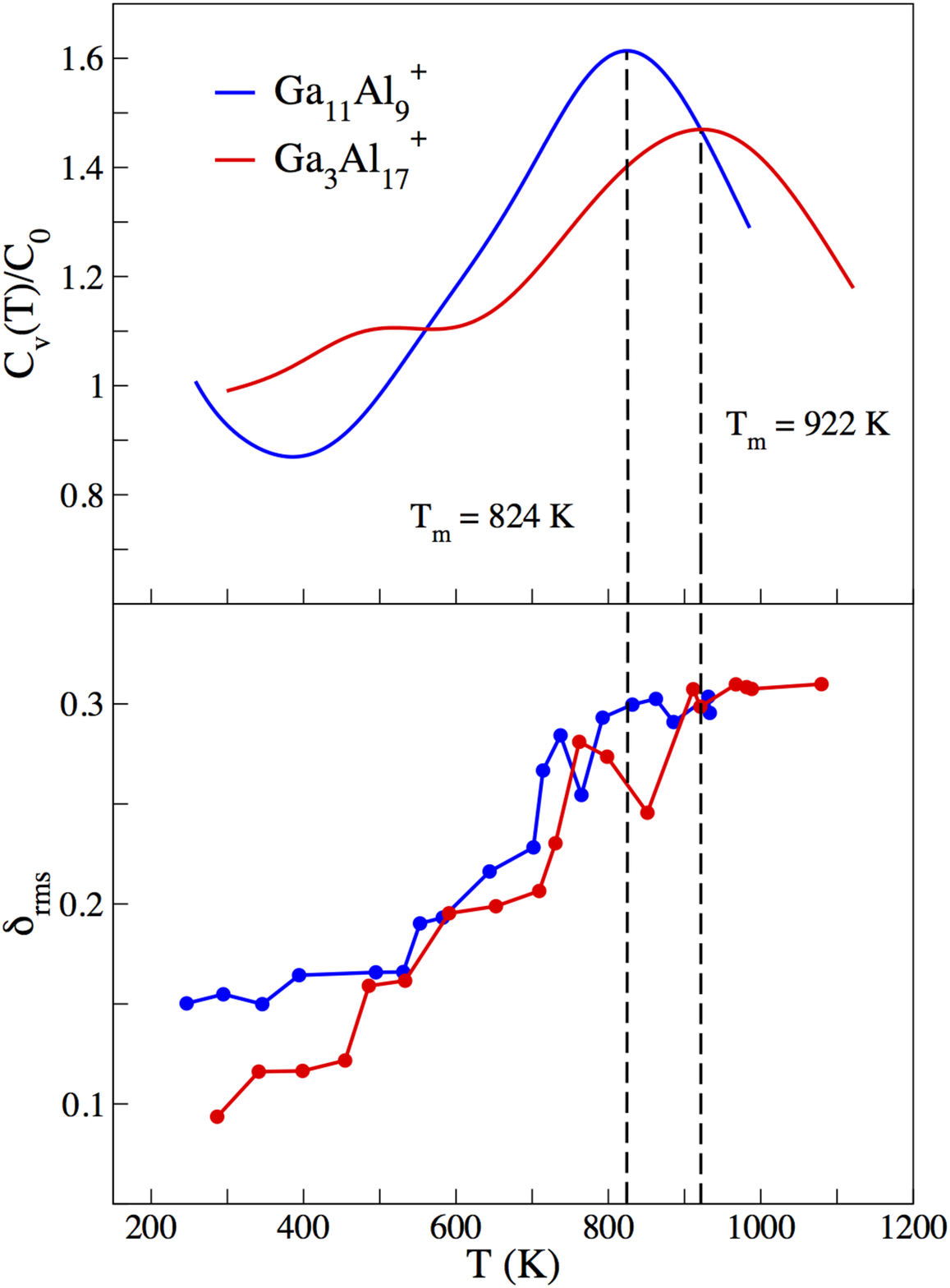}
\caption{Comparison of canonical specific heat curves of Ga\textsubscript{11}Al\textsubscript{9}\textsuperscript{+} and Ga\textsubscript{3}Al\textsubscript{17}\textsuperscript{+} (top panel) and the root-mean-squared (RMS) bond-length fluctuations (bottom panel) as a function of temperature.}
\label{spheat}
\end{figure}

The top panel of Fig.~\ref{spheat} shows the canonical heat capacity curves obtained from the MH method for Ga\textsubscript{11}Al\textsubscript{9}\textsuperscript{+} and Ga\textsubscript{3}Al\textsubscript{17}\textsuperscript{+} clusters normalized to the classical specific heat capacity ($C_{0} = (3N - 6 + 3/2)k_{B}$ where N is the number of atoms and $k_{B}$ is the Boltzmann constant) accounting for rotational and vibrational contributions. The solid-liquid-like phase transition temperature, corresponding to the peak in canonical specific heat capacity plot, is 824 K for Ga\textsubscript{11}Al\textsubscript{9}\textsuperscript{+} and 922 K for Ga\textsubscript{3}Al\textsubscript{17}\textsuperscript{+} cluster respectively. 

Analyses of the indices pertaining to the cluster structure at different temperatures help to relate the structural changes occurring in the cluster to its thermodynamic behaviour. Melting, in the context of clusters, is a change from the \textit{solid-like} state to the \textit{liquid-like} state. The root mean squared (RMS) bond-length fluctuation (Berry parameter\cite{Berry_parameter}) identifies this phase transition i.e. the melting-like temperature as the temperature above which the average change in bond-lengths converge. Expressed as $\delta_{rms}$, 

\begin{equation}
\delta_{rms} = \frac{2}{N\cdot(N-1)}\sum_{i>j}\frac{(<r_{ij}^{2}>_{t} - <r_{ij}>_{t}^{2})^{1/2}}{<r_{ij}>_{t}}
\end{equation}
\\
where N is the total number of atoms, $r_{ij}$ is distance between atoms i and j and $<>_{t}$ denotes the time average, this index of fluctuation quantifies the response of bond vibrations to increasing energy (or temperature). 

Shown in the bottom panel of Fig.~\ref{spheat} are the calculated $\delta_{rms}$ values for Ga\textsubscript{11}Al\textsubscript{9}\textsuperscript{+} (shown in blue) and Ga\textsubscript{3}Al\textsubscript{17}\textsuperscript{+} (shown in red) clusters. The fluctuations in bond-length are quite low $(\sim 0.1- 0.15)$ at the lowest energy and start to increase with temperature. However, above the calculated solid-liquid-like phase transition temperature of 824 K in Ga\textsubscript{11}Al\textsubscript{9}\textsuperscript{+} and 922 K in Ga\textsubscript{3}Al\textsubscript{17}\textsuperscript{+}, the values converge to $(\sim 0.3)$ indicating a structural transition. It must, however, be stated that although at low and high temperatures the values of $\delta_{rms}$ are consistent with PT independent simulations, the absolute value of the index in the intermediate temperature range is subject to the influence of PT.  

\begin{figure}[t!]
\centering
\includegraphics[trim = 0cm 0cm 0cm 0cm,clip=true, width=\columnwidth]{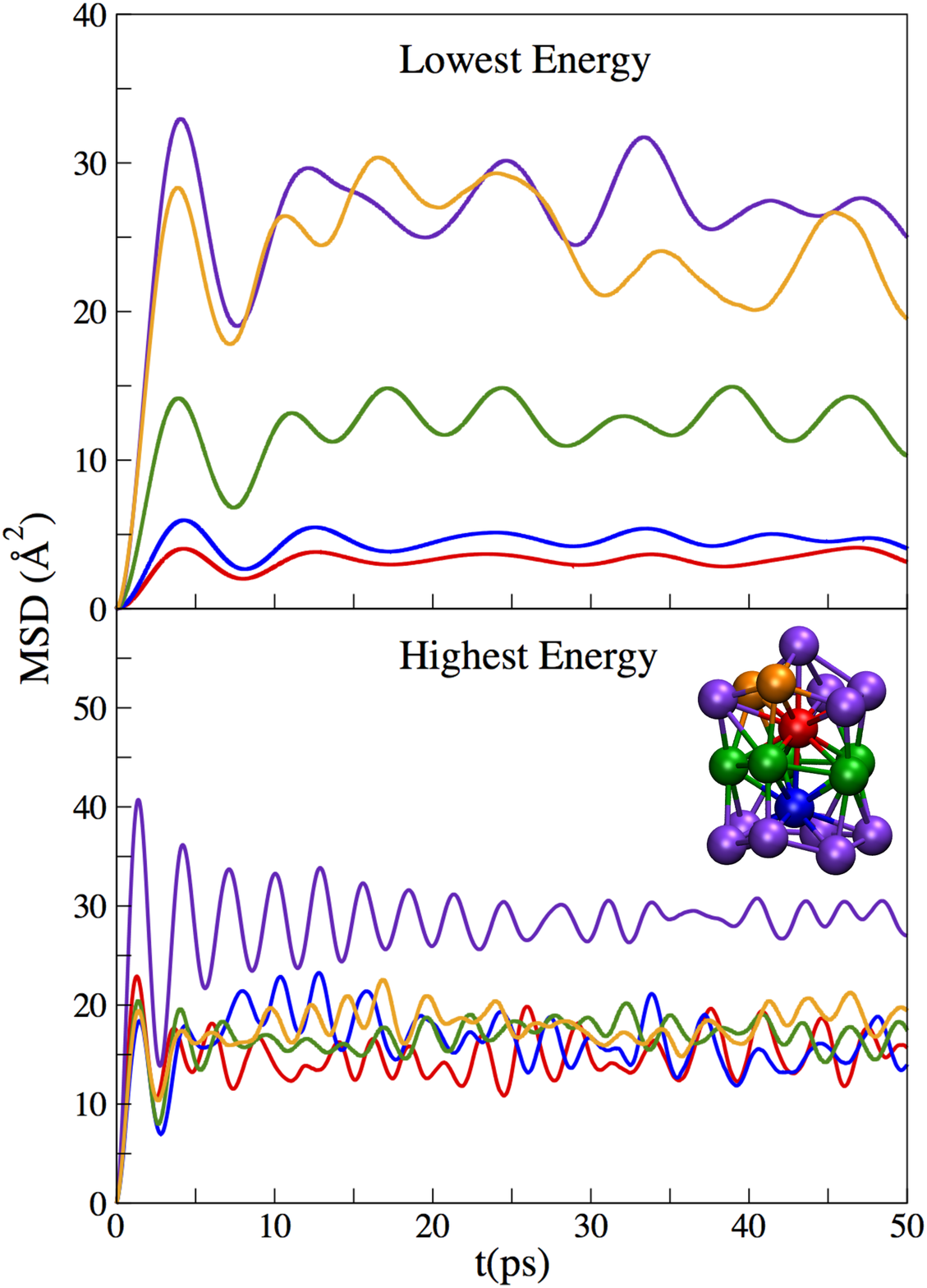}
\caption{Comparison of mean squared displacement (MSD) of Ga\textsubscript{11}Al\textsubscript{9}\textsuperscript{+} at lowest (246 K) and highest energy (931 K) as a function of time. Atoms in red and blue are the central Al atoms; green and orange corresponds to MSD averaged over all the Al atoms in the central and top ring respectively; indigo represents the average MSD of Ga atoms as also shown in inset. For references in the text, the more highly coordinated internal atom (red) is referred as \textit{a} and the other (coloured blue) as \textit{b}.}
\label{msdGa11Al9}
\end{figure}

Mean squared displacement (MSD) helps to compare the relative mobilities of the internal and surface atoms. Calculated as per Eq.~\ref{msd}, 
\begin{equation}
<r_{i}(t)> = \frac{1}{M}\sum_{m=1}^{M}[R_{i}(t_{m}+t) - R_{i}(t_{m})]^{2}
\label{msd}
\end{equation}
where M is the number of time origins, the individual atomic contributions to the overall behavior of the system can be analyzed. Fig.~\ref{msdGa11Al9} shows the MSD of gallium and aluminium atoms of Ga\textsubscript{11}Al\textsubscript{9}\textsuperscript{+} at the lowest and highest energy. Coloured in indigo, the amplitude of the MSD of the surface gallium atoms (averaged) does not change significantly as the energy is increased. Al atoms present at the internal sites (red and blue) show the maximum increase in the amplitude of MSD from lowest to the highest energy. Moreover, at the highest energy all the Al atoms have almost similar MSDs indicating a \textit{liquid-like} phase.     

\begin{figure}[t!]
\centering
\includegraphics[trim = 0cm 0cm 0cm 0cm,clip=true, width=\columnwidth]{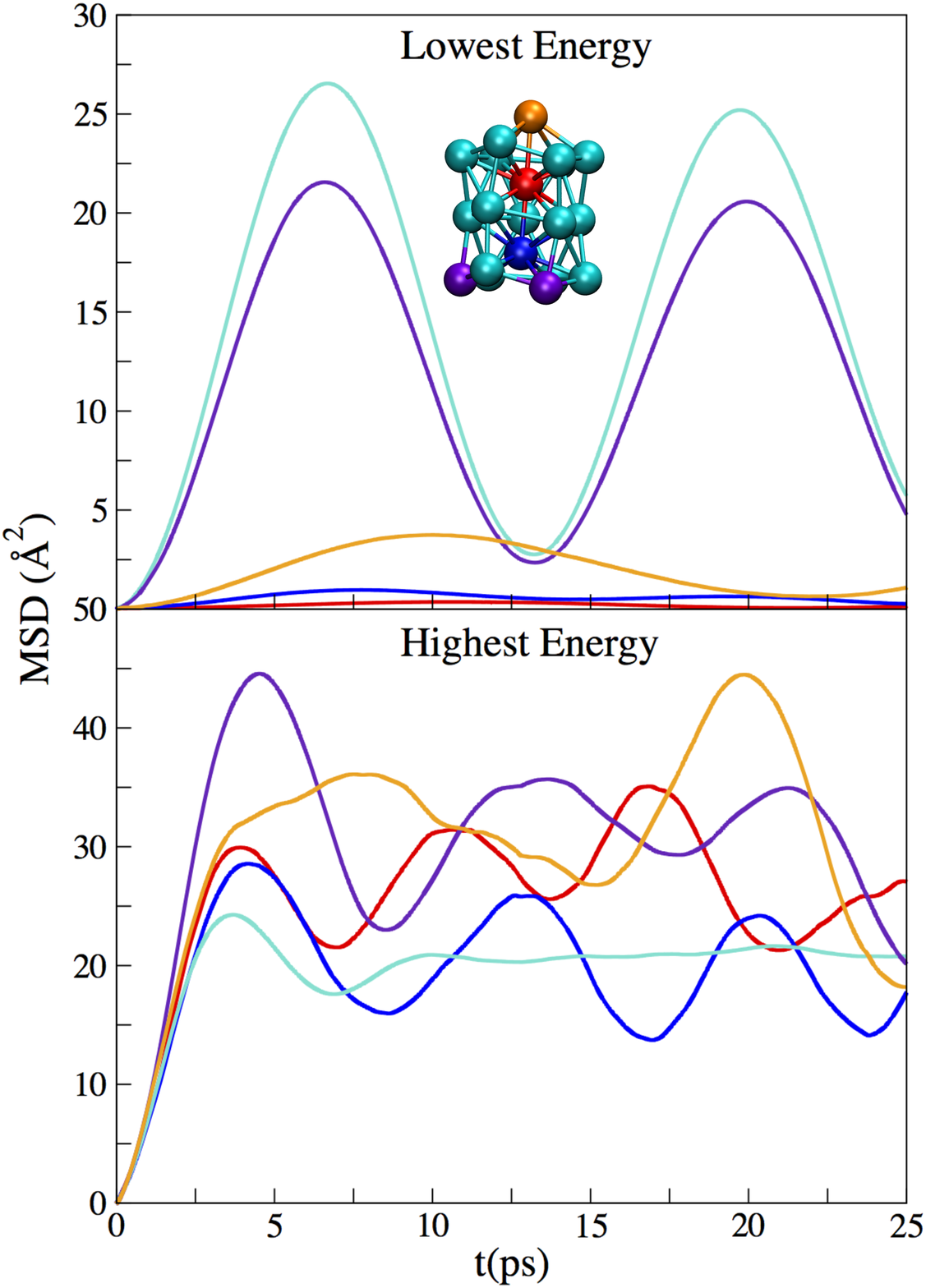}
\caption{Comparison of mean squared displacement (MSD) of Ga\textsubscript{3}Al\textsubscript{17}\textsuperscript{+} at lowest (286 K) and highest energy (1079 K) as a function of time. Atoms in red and blue are the central Al atoms; turquoise and indigo corresponds to MSD averaged over all the surface Al and bottom ring Ga atoms respectively. Orange represents the MSD of top Ga atom. For references in the text, the more highly coordinated internal atom (red) is referred as \textit{a} and the other (coloured blue) as \textit{b}.}
\label{msdGa3Al17}
\end{figure}
Fig.~\ref{msdGa3Al17} shows the MSD calculated for Ga\textsubscript{3}Al\textsubscript{17}\textsuperscript{+} at the lowest and highest energy. In contrast to the Ga atoms in Ga\textsubscript{11}Al\textsubscript{9}\textsuperscript{+} (average MSD of all Ga atoms coloured indigo in Fig.~\ref{msdGa11Al9}) where the MSD have similar amplitudes for all the atoms at the lowest and highest energies, one of the three Ga atoms situated at the top (colored orange in inset) has MSD values between those of internal atoms and other surface atoms. This difference, however, vanishes as the energy is increased and all the surface atoms have similar MSD values. It is for the internal Al atoms (red and blue) that the MSD changes with increasing temperature. At the highest energy, all the atoms (Ga and Al) have MSD values reflecting a \textit{liquid-like} state in Ga\textsubscript{3}Al\textsubscript{17}\textsuperscript{+}.        

The diffusion coefficient helps to quantify the mobility of atoms in the cluster. It is calculated as per Eq.~\ref{diffcoeff} where D is the diffusion coefficient of the specific atom, $<r_{i}(t)>$ is the mean squared displacement of the corresponding atom calculated as per Eq.~\ref{msd} and d is the dimensionality of the cluster system (which is 3). For small systems such as clusters, the diffusion coefficient is meaningful only over a small time range where the MSD achieves a constant slope\cite{berry_msd}.  
\begin{equation}
D \equiv \frac{1}{2d}\lim_{t\rightarrow\infty}\frac{<r_{i}(t)>}{t}
\label{diffcoeff}
\end{equation}

\begin{figure}[t]
\centering
\includegraphics[trim = 0cm 0cm 0cm 0cm,clip=true, width=\columnwidth]{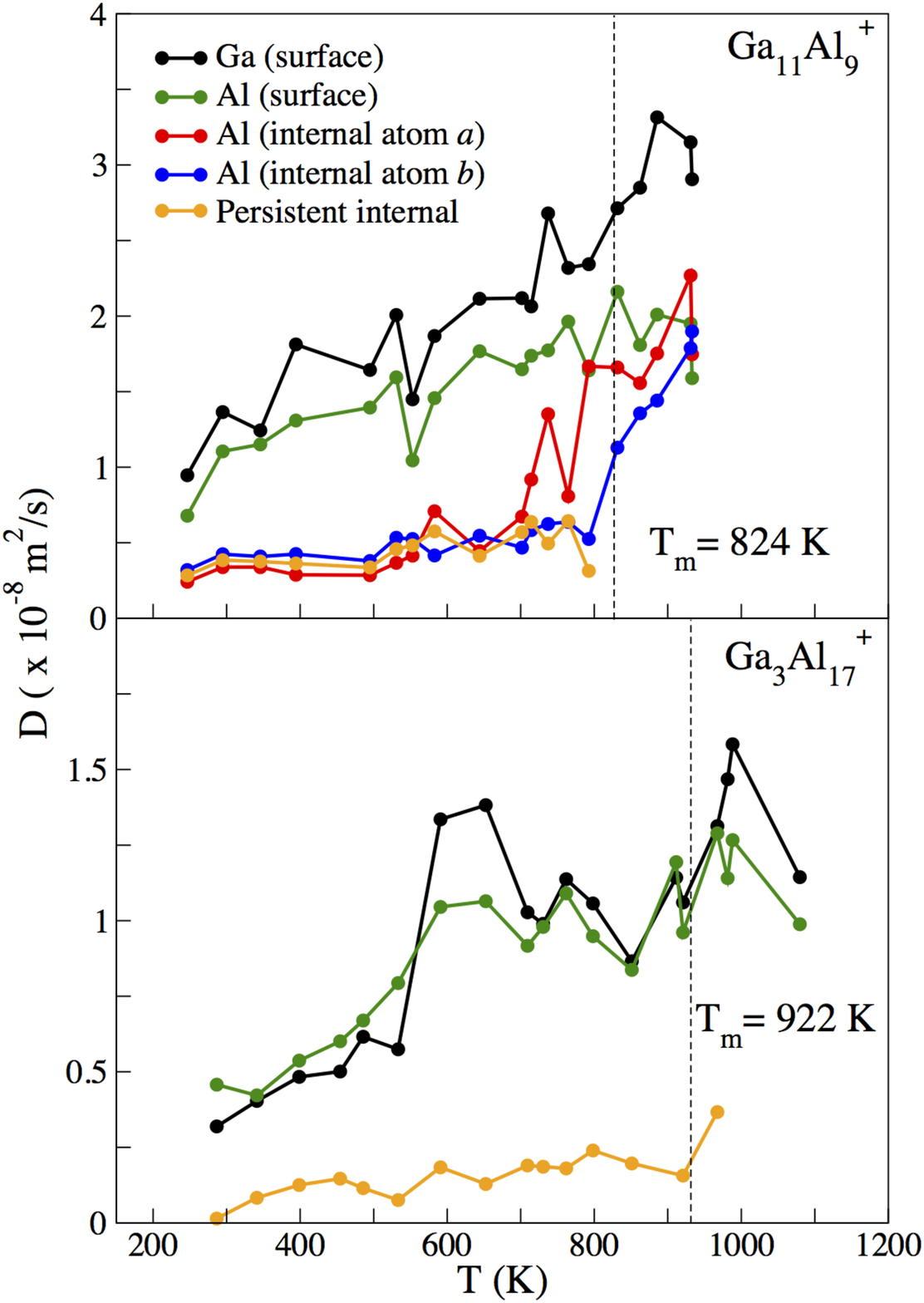}
\caption{Diffusion coefficient of internal and surface atoms as a function of temperature. The vertical dotted line denotes the melting temperature.}
\label{DC}
\end{figure}

Shown in Fig.~\ref{DC} are diffusion coefficient values calculated for Ga\textsubscript{11}Al\textsubscript{9}\textsuperscript{+} and Ga\textsubscript{3}Al\textsubscript{17}\textsuperscript{+} clusters in the top and bottom panels respectively. The MSD values over which the slope is calculated have been kept fixed between 1 and 2 \AA$^2$. MSD values of all the surface gallium and aluminium atoms have been averaged as shown in black and green for both the clusters. In both cases, the diffusion coefficients of the surface atoms are very high with the trend increasing with temperature. The average coordination number, calculated using the first minimum in pair distribution function $g(r)$ as cutoff distance for coordination, of Ga\textsubscript{11}Al\textsubscript{9}\textsuperscript{+} (Fig.~1, Supplementary Information) shows that the internal atoms (red and blue) occupy an internal site at all temperatures below the melting temperature of 824 K. In the top panel of Fig.~\ref{DC}, the diffusion coefficient of both the internal atoms is much lower at all temperatures up to 701 K after which the internal atom \textit{a} (coloured red in inset of bottom panel, Fig.~\ref{msdGa11Al9}) shows fluctuations in Ga\textsubscript{11}Al\textsubscript{9}\textsuperscript{+}. However, it is only after the melting temperature that the diffusion coefficient of the internal atoms shoots up and at the highest average temperature becomes comparable with the rest of the Al atoms in the Ga\textsubscript{11}Al\textsubscript{9}\textsuperscript{+} cluster. 

Although the surface Ga and Al atoms of Ga\textsubscript{3}Al\textsubscript{17}\textsuperscript{+} show similar behaviour with increasing temperature,  constant exchange among atoms to occupy an internal site at temperatures closer to the melting temperature (as also clear from the average coordination number plot in Fig.~2, Supplementary Information) makes it much harder to tag a particular atom as \textit{internal}. To counter this situation, we attribute the term `\textit{persistent internal}' to an atom which satisfies the following two conditions: (a) the average coordination number has to be greater-than-or-equal-to 9.5 and, (b) the time during which condition (a) is satisfied during the continuous MD run has to be more than 1.5 ps. The second condition also helps to make sure that an atom occupying an internal site at a particular energy run is favourable rather than being an artefact of parallel tempering.   
 
Coloured orange in the top and bottom panels of Fig.~\ref{DC}, the diffusion coefficient of the `\textit{persistent internal}' atom(s) in both the clusters is much lower in comparison to the surface atoms below the melting temperature. Above the melting temperature, none of the atoms in Ga\textsubscript{11}Al\textsubscript{9}\textsuperscript{+} satisfy the condition to be considered as `\textit{persistent internal}'. In Ga\textsubscript{3}Al\textsubscript{17}\textsuperscript{+}, in contrast, the `\textit{persistent internal}' atom criteria is fulfilled up to 967 K, above the melting temperature of 922 K. Moreover, in Ga\textsubscript{11}Al\textsubscript{9}\textsuperscript{+} the diffusion coefficient of the `\textit{persistent internal}' atom is an average of both the internal atoms (red and blue) up to 582 K. 


The Quantum Theory of Atoms in Molecules (QTAIM)\cite{bader} aims to define atomic boundaries within a molecule using the topology of the total electronic charge density of the molecular system. The volume occupied by an atom, which is an open system, is governed by the topological condition of zero-flux of the electronic charge density ($\bigtriangledown \rho ( r )\cdot n ( r ) = 0$). Integrating the electron density within each atomic volume gives the total electronic charge associated with the corresponding ion. 

\begin{figure}[t]
\centering
\subfigure[Ga$_{20}^{+}$]{\includegraphics[trim = 1.7cm 1cm 1.7cm 0cm,clip=true,width=0.325\columnwidth]{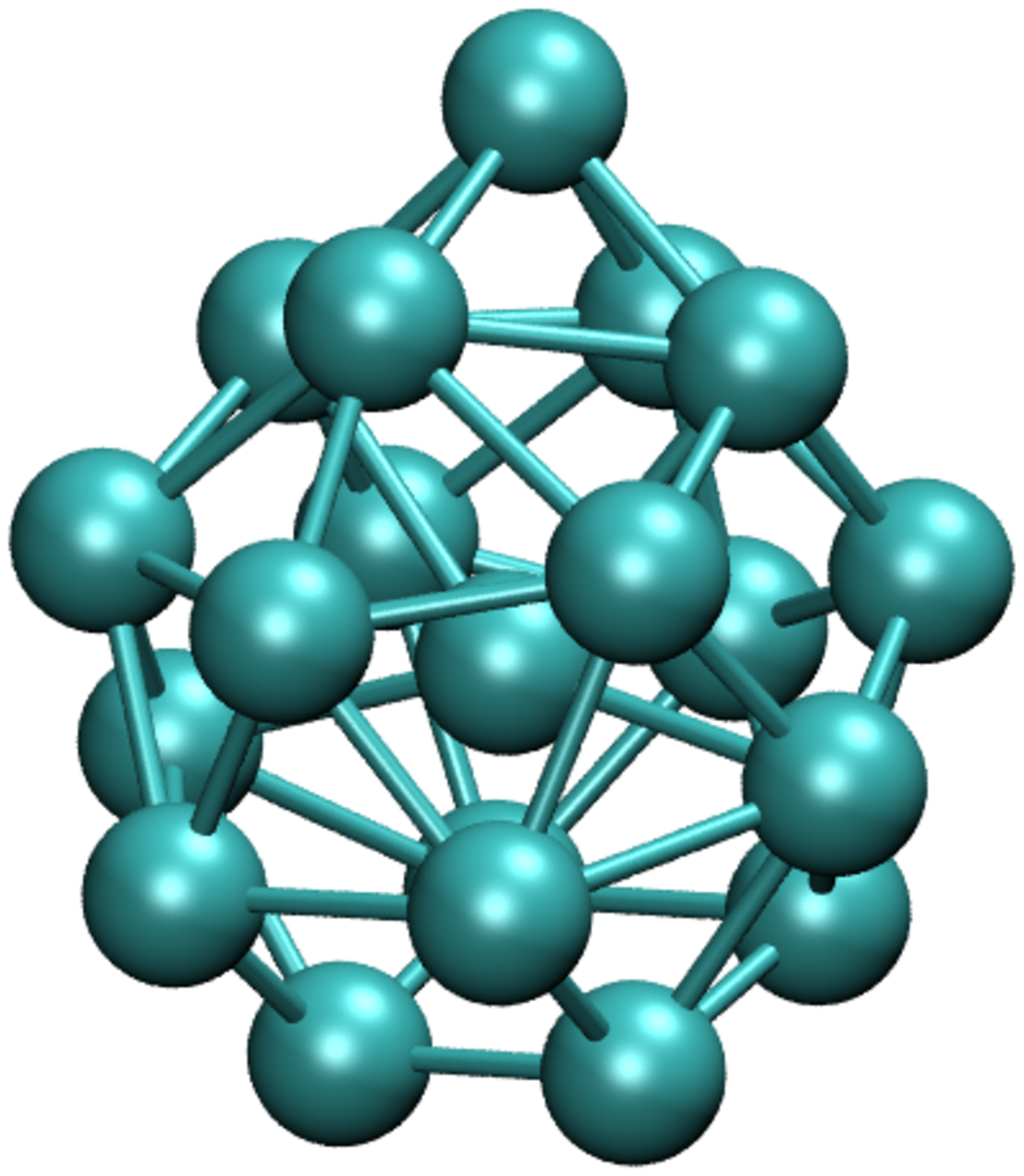}}
\subfigure[Ga\textsubscript{19}Al\textsuperscript{+}]{\includegraphics[trim = 1.7cm 1cm 1.7cm 0cm,clip=true,width=0.325\columnwidth]{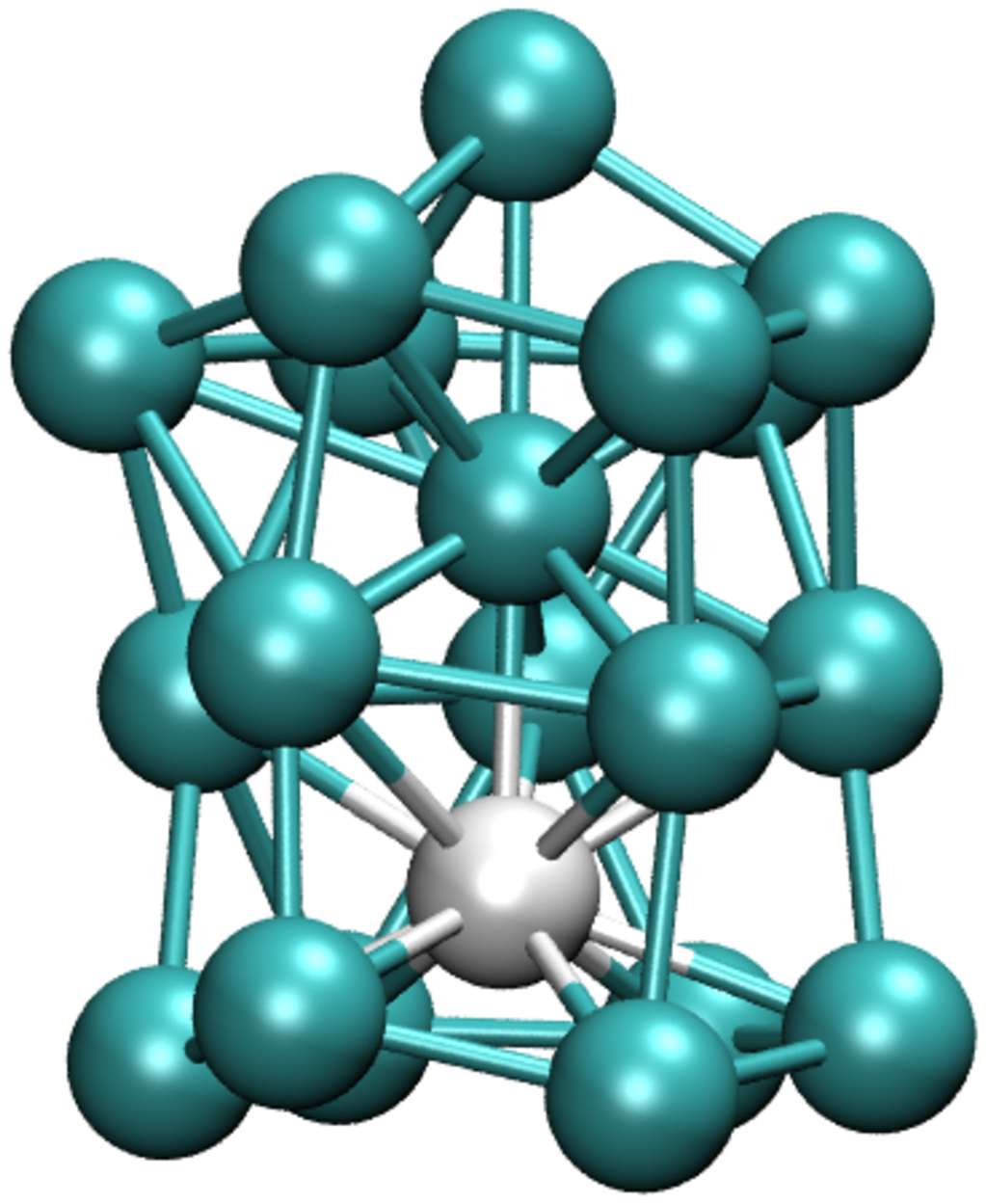}}
\subfigure[Ga\textsubscript{11}Al\textsubscript{9}\textsuperscript{+}]{\includegraphics[trim = 1.7cm 1cm 1.7cm 0cm,clip=true,width=0.325\columnwidth]{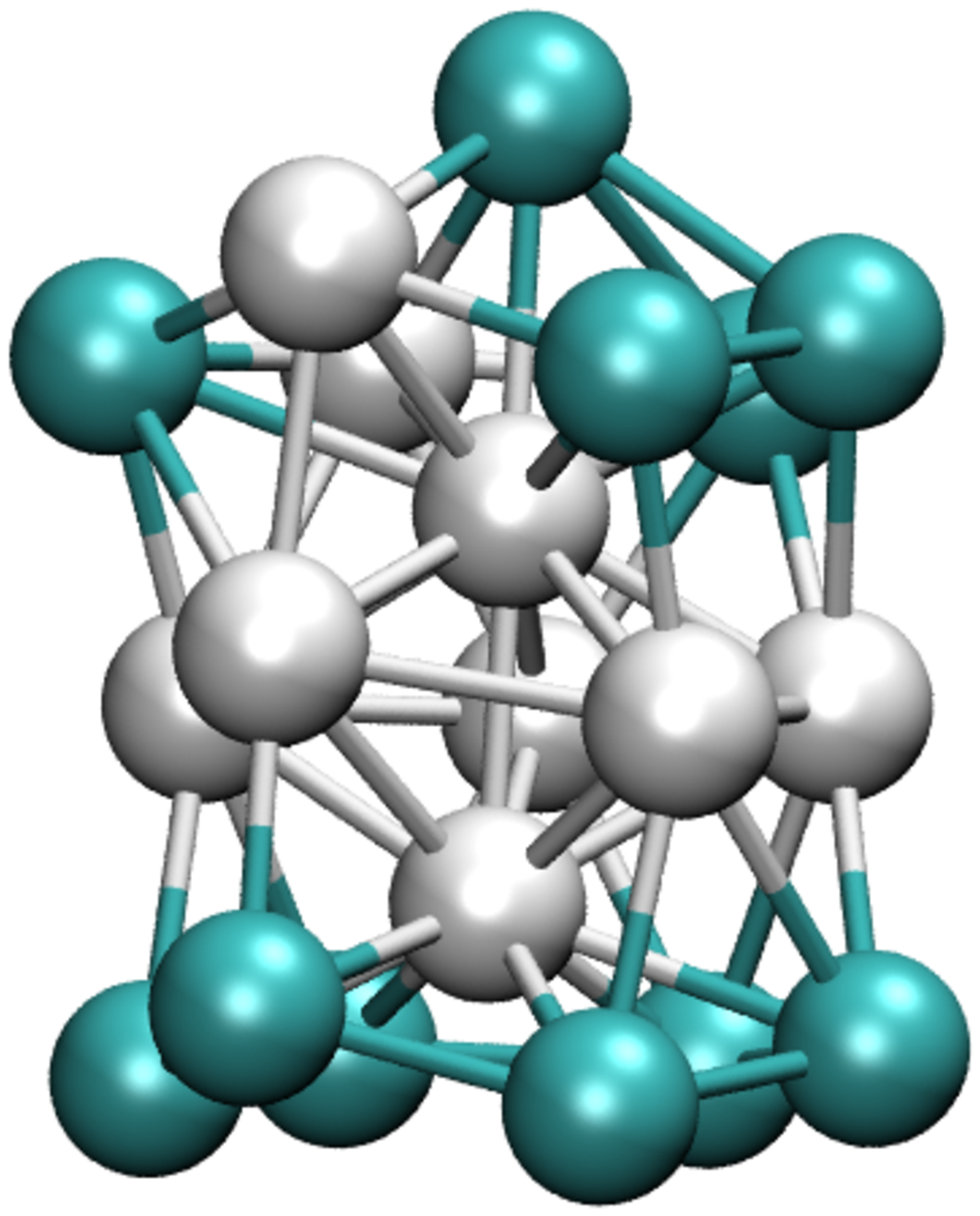}}
\subfigure[Ga\textsubscript{3}Al\textsubscript{17}\textsuperscript{+}]{\includegraphics[trim = 1.7cm 1cm 1.7cm 0cm,clip=true,width=0.325\columnwidth]{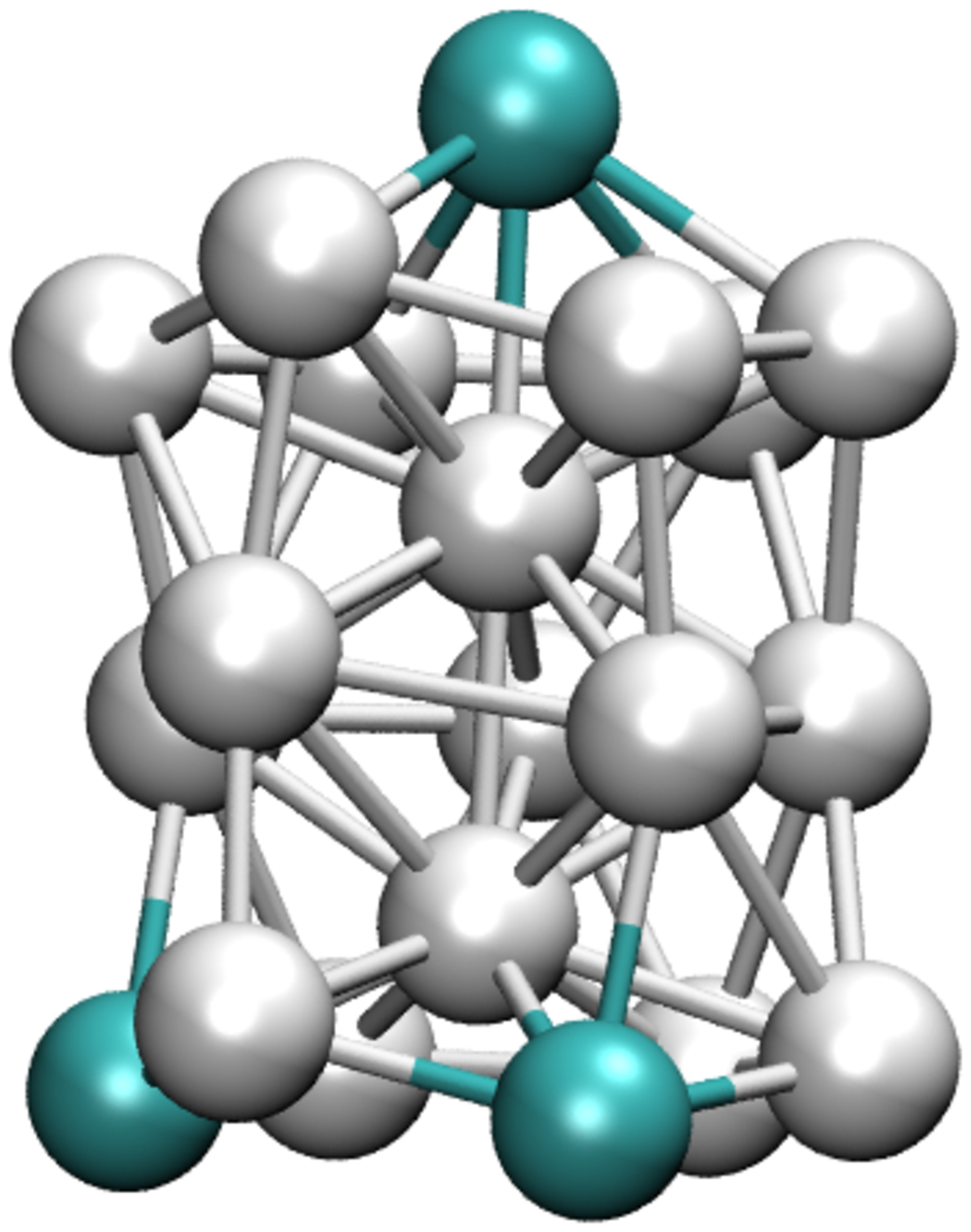}}
\subfigure[Al\textsubscript{20}\textsuperscript{+}]{\includegraphics[trim = 1.7cm 1cm 1.7cm 0cm,clip=true,width=0.325\columnwidth]{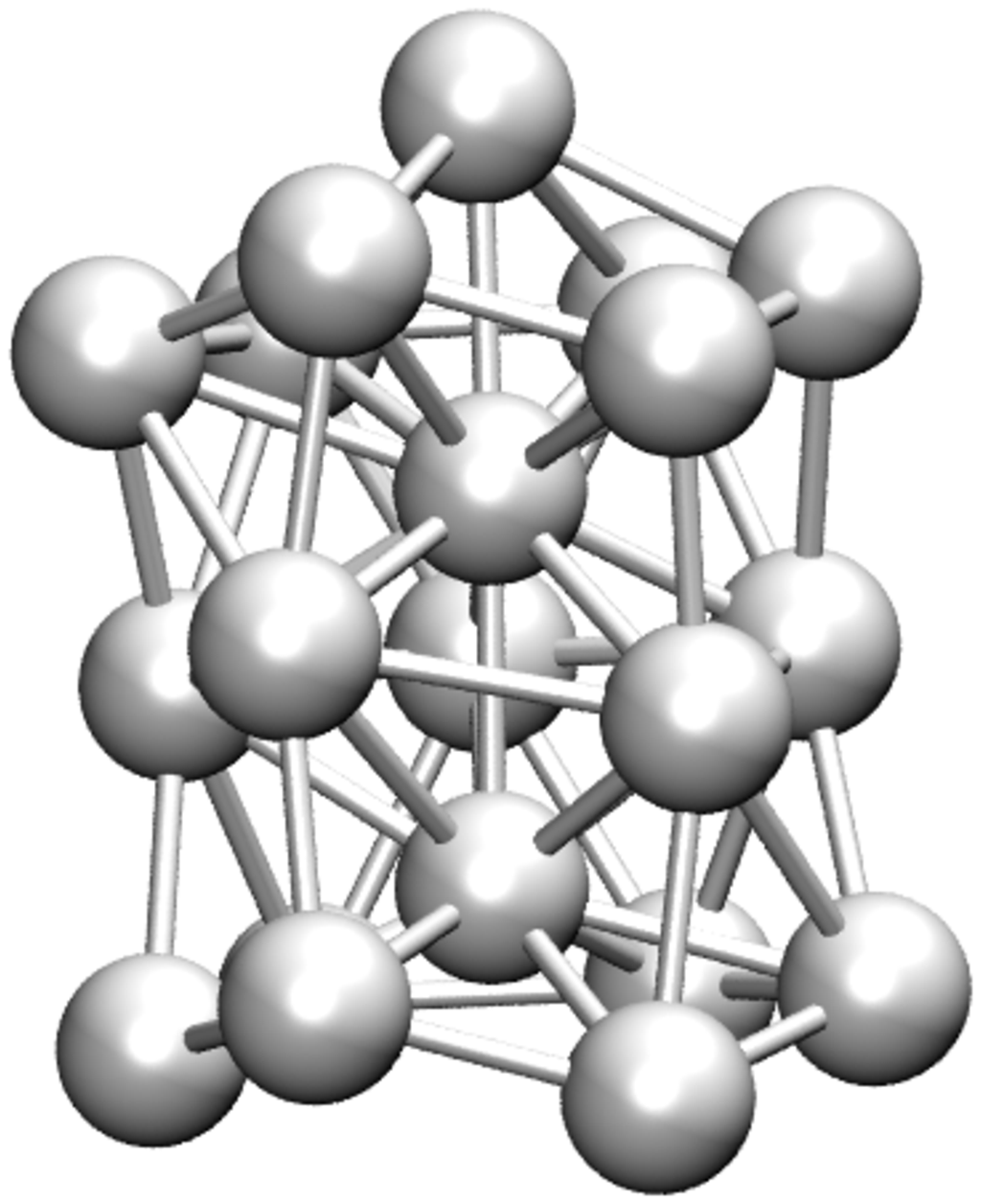}}
\caption{ Starting structures in the capped sphere (Ga$_{20}^{+}$) and stacked plane (Ga\textsubscript{19}Al\textsuperscript{+}, Ga\textsubscript{11}Al\textsubscript{9}\textsuperscript{+}, Ga\textsubscript{3}Al\textsubscript{17}\textsuperscript{+} and Al\textsubscript{20}\textsuperscript{+}) configurations with Al atoms in white and Ga atoms in cyan.}
\label{structure}
\end{figure}

Shown in Fig.~3 (Supplementary Information) are the partial charges associated with each of the gallium and aluminium atoms in Ga\textsubscript{11}Al\textsubscript{9}\textsuperscript{+} and Ga\textsubscript{3}Al\textsubscript{17}\textsuperscript{+} respectively. There is a strong charge segregation between the internal and surface atoms in both the cases. Moreover, among the charged surface atoms it is the gallium atom which becomes negatively charged and aluminium which becomes positively charged in both cases. A similar picture, where an electrostatic cage confines the internal atoms was also observed for the monometallic Ga\textsubscript{20}\textsuperscript{+} and Al\textsubscript{20}\textsuperscript{+} clusters\cite{selfAl}.
\subsection{Phase diagram of Ga-Al}

Incorporating the results from our previous investigations\cite{self,selfAl} of Ga\textsubscript{20}\textsuperscript{+}, Al\textsubscript{20}\textsuperscript{+} and Ga\textsubscript{19}Al\textsuperscript{+}, we have created a binary phase diagram of the 20 atom Ga-Al nanocluster system. Irrespective of the bimetallic cluster's composition, the solid-liquid-like phase transition temperature is always above the corresponding bulk alloy melting temperature\cite{bulkGaAl} as also shown in Table~\ref{melttemp}.

\begin{table}[b]
\centering
\caption{Comparison of the melting temperatures of Ga-Al alloys in cluster and bulk phase\cite{bulkGaAl}.}
\begin{tabular}{c c c c}
\hline
\hline
 & Cluster T$_{m}$(K) & Bulk T$_{m}$(K) & $\Delta T_m$(K) \\
 \hline
Ga$_{20}^{+}$ & 616 & 303 & 313 \\
Ga\textsubscript{19}Al\textsuperscript{+} & 686 & 353 & 333 \\
Ga\textsubscript{11}Al\textsubscript{9}\textsuperscript{+} & 824 & 646 & 178 \\
Ga\textsubscript{3}Al\textsubscript{17}\textsuperscript{+} &  922 & 857 & 65 \\
Al\textsubscript{20}\textsuperscript{+} & 993 & 933 & 60 \\
\hline
\end{tabular}
\label{melttemp}
\end{table}

\begin{figure}[t!]
\centering
\includegraphics[trim = 0cm 0cm 0cm 0cm,clip=true,width=\columnwidth]{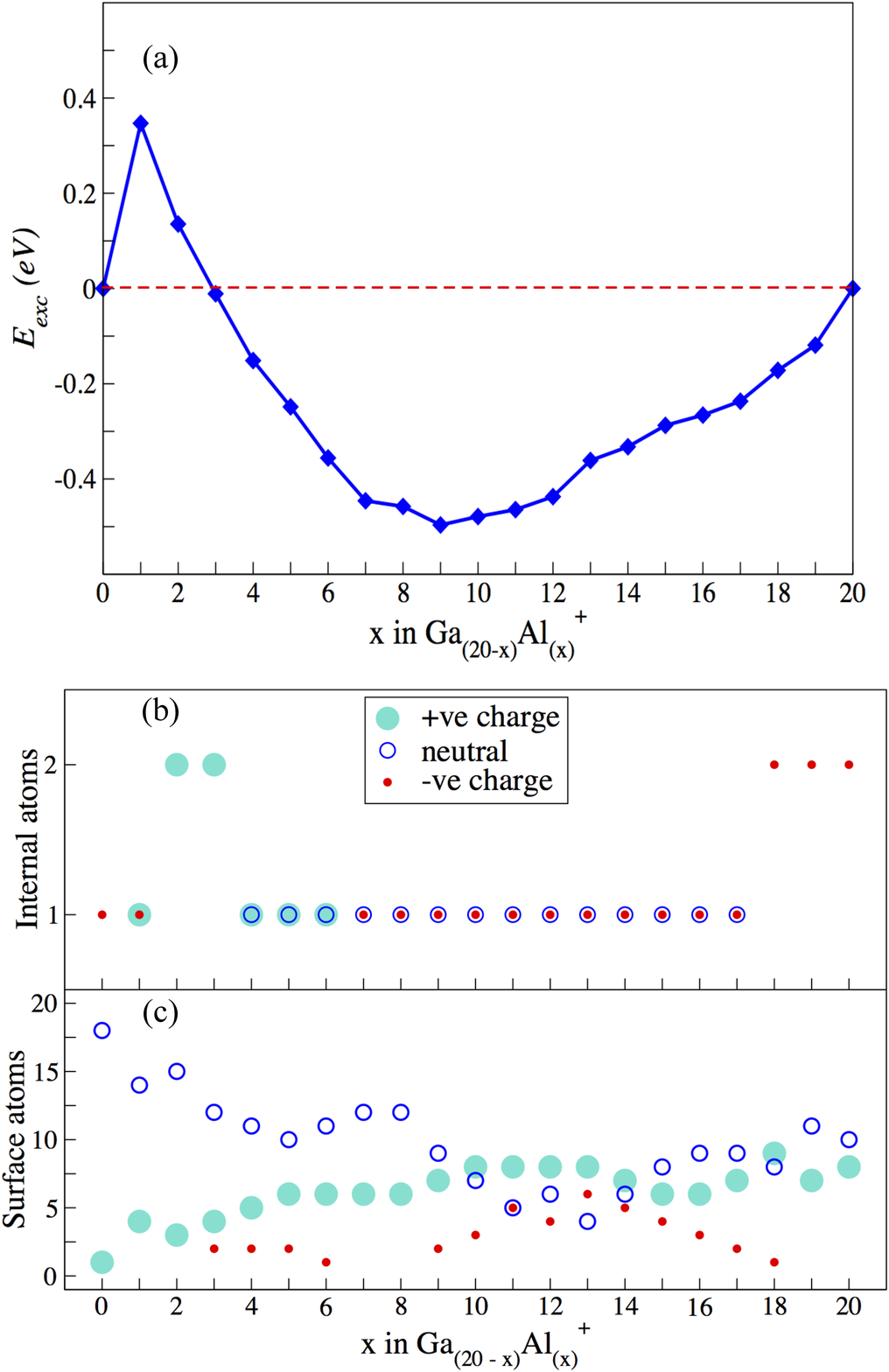}
\caption{Excess energy for all the compositions in Ga$_{20-x}$Al$_{x}$$^+$, panel (a), and the corresponding atoms-in-molecules analysis where the red dot, blue open circle and turquoise circle represents negative, neutral and positive charges.}
\label{mixingenergy}
\end{figure}

The energy of mixing, calculated as per Eq.~\ref{excessenergy}, reflects the energy gain (or loss) for a mixed cluster relative to the pure clusters. A negative value indicates mixing i.e. formation of the nanoalloy is energetically favourable, whereas a positive value indicates segregation between the participating atoms.      

\begin{equation}
E_{exc} = E_{tot}^{Ga_{(20-x)}Al_{x}^{+}} - \frac{x}{20}E_{tot}^{Al_{20}^{+}} - \frac{20-x}{20}E_{tot}^{Ga_{20}^{+}}
\label{excessenergy}
\end{equation} 
Fig.~\ref{mixingenergy}~(a) shows the energy of mixing for all the cluster compositions in the series Ga$_{(20-x)}$Al$_{x}$$^+$. Adding an aluminium atom to Ga$_{20}^{+}$ gives a positive value to the energy of mixing indicating segregation of the Al atom in the gallium dominated Ga\textsubscript{19}Al\textsuperscript{+} cluster. The energy of mixing remains positive for Ga$_{18}$Al$_{2}$$^+$ and from Ga$_{17}$Al$_{3}$$^+$ onwards it becomes negative, achieving a minimum at the most favourable composition, i.e. Ga\textsubscript{11}Al\textsubscript{9}\textsuperscript{+}. This trend is also visible from the change in melting temperature between the Ga$_{(20-x)}$Al$_{x}$$^+$ clusters and the corresponding bulk alloy counterparts as shown in Table\ref{melttemp}. The change in melting temperature first increases from Ga\textsubscript{20}\textsuperscript{+} to Ga\textsubscript{19}Al\textsuperscript{+} but thereafter decreases with increasing Al composition thus reflecting how the energy of mixing affects the greater-than-bulk melting behaviour of Ga\textsubscript{(20-x)}Al\textsubscript{x}\textsuperscript{+} clusters.

In order to understand further how the charge distribution among atoms, i.e. with respect to internal \& external sites, affects the segregation or mixing behaviour in each cluster composition in Ga$_{(20-x)}$Al$_{x}$$^+$, we performed atoms-in-molecules analysis as shown in panel (b) and (c) of Fig.~\ref{mixingenergy}. The optimized structure for each of the nanoalloy clusters was obtained from the SP configuration of Ga$_{20}$ as has been explained in Section 2. From the obtained partial charges on each atom ($q_i$), an atom is considered neutral if $-0.1e \le q_{i} \le 0.1e$, positively charged if $q_{i} > 0.1e$ and negatively charged if $q_{i} < -0.1e$. The Ga$_{20}^{+}$ capped sphere structure\cite{krista1} has one internal negatively charged atom and one positively charged surface atom with the remaining atoms being neutral. In Ga\textsubscript{19}Al\textsuperscript{+}, the internal Al and Ga atom becomes strongly positively and negatively charged respectively, thus forming a dipole with no negatively charged surface atoms. In Ga$_{18}$Al$_2$$^+$ and Ga$_{17}$Al$_3$$^+$, both the internal atoms are positively charged. However, for Ga$_{17}$Al$_3$$^+$, there is a negative charge on the surface. Starting from Ga$_{16}$Al$_4$$^+$, there is one neutral Al atom occupying the internal site with the other being positively charged until Ga$_{14}$Al$_6$$^+$ and negatively charged from Ga$_{13}$Al$_7$$^+$ to Ga$_{17}$Al$_3$$^+$. Both the internal atoms are negatively charged for Ga$_{2}$Al$_{18}$$^+$, GaAl$_{19}$$^+$ and Al\textsubscript{20}\textsuperscript{+}. Correlating with the energy of mixing, as shown in Fig.~\ref{mixingenergy}(a), we see that segregation (a positive energy of mixing) corresponds to either the formation of a strong dipole between the internal atoms (as in Ga\textsubscript{19}Al\textsuperscript{+}) or the absence of negative charge at the centre or surface (as in Ga$_{18}$Al$_2$$^+$). Moreover, for Ga$_{13}$Al$_7$$^+$, Ga$_{12}$Al$_8$$^+$ and GaAl$_{19}$$^+$, along with the monometallic Ga\textsubscript{20}\textsuperscript{+} and Al\textsubscript{20}\textsuperscript{+} clusters, there is no negatively charged surface atom. Strong charge segregation between the internal and surface sites plays an important role in the greater-than-bulk melting behaviour as also reflected by the diffusion coefficients (Fig.~\ref{DC}) of the representative clusters\cite{selfAl}. 

The average coordination number, diffusion coefficient and atoms-in-molecules analysis bring out the differences between the internal and surface sites in  these clusters. The internal atoms being confined in an electrostatic cage, breaking of which leads to the solid-liquid-like phase transition in Ga\textsubscript{(20-x)}Al\textsubscript{x}\textsuperscript{+} clusters, is a physical picture which can help explain the \textit{greater-than-bulk melting character} (GB) of this system. Greater-than-bulk melting behaviour, observed in gallium\cite{jarrold_GaPRL} and tin\cite{jarrold_tin} clusters, is an exception to the melting point depression with size as predicted by Pawlow\cite{pawlow}. Models typically applied to understand the size dependence of the phase transition in clusters\cite{pd4,bulk2} have expressed the cluster melting temperature as a function of the bulk melting temperature with a parameter catering for the size dependence component (where the nanoparticle is generally assumed spherical). Although this may be a reasonable approximation for cluster sizes having hundreds of atoms, a plethora of research currently addresses the different spatial geometries assumed by the putative global minimum structures of nanoparticles having fewer than tens of atoms\cite{baletto,nanoalloys}.   

How different is a gallium-aluminium nanoalloy cluster from the corresponding bulk alloy material? Taking into account the GB character, energy of mixing and the associated intrinsic elemental character (to be common in bulk and cluster), we have attempted to provide a simplistic view of the Ga\textsubscript{(20-x)}Al\textsubscript{x}\textsuperscript{+} cluster system. Our contention is that a cluster can be viewed as a piece of the bulk matter with additional properties specific to a particular cluster size, as expressed in Eq.~\ref{model} for our specific case. Our intention is two-fold: firstly, to find a way to `quantify' the inherent and size dependent GB character among Ga\textsubscript{(20-x)}Al\textsubscript{x}\textsuperscript{+} clusters and, secondly to compare the obtained melting temperature from Eq.~\ref{model} with those obtained from our DFT calculations to gauge the applicability of this picture.    
\begin{multline}
T_{m} (Ga_{(20-x)}Al_{x}^{+}) = T_{bulk} (Ga:Al :: (20-x):x) \\
	+ \nu_{GB} + \nu_{exc}
\label{model}
\end{multline}
Here T\textsubscript{m} and T\textsubscript{bulk} are the cluster and bulk alloy melting temperatures for similar composition and $\nu_{GB}$, $\nu_{exc}$ are the functions representing the contributions of the greater-than-bulk melting character and the energy of mixing respectively.   

To `quantify' the GB character, we revert back to the picture of the internal atoms which are confined in an electrostatic cage by the surface atoms in Ga\textsubscript{(20-x)}Al\textsubscript{x}\textsuperscript{+} clusters, disruption of which is required for melting. We view this difference in the environments of the internal and surface atoms as an `energy barrier' which exists between the internal and surface sites. With increasing temperature, the internal and surface atoms have enough energy to cross this existing `energy barrier' and thus either replace each other or independently become a surface or internal atom which subsequently leads to a change in their respective mobilities in their new sites below the melting temperature. This change in the cluster either due to a change in the number of internal atoms or the type of internal atom (considering swapping of atoms due to parallel tempering or an exchange of positions between the internal and surface atoms) is, from here onwards, termed as \textit{`Change-Swap (CS)'}. Characterising the \textit{liquid-like} state as one where all the atoms have an equal likelihood of occupying an internal or surface site, we have attempted to understand the strength of the `energy barrier' among the five representative structures in Ga$_{(20-x)}$Al$_{x}$$^+$ series through an \textit{Arrhenius-like} equation, Eq.~\ref{Arrh}, which is also used to describe diffusion. 
\begin{equation}
ln~k_{CS} = ln~A  - {E_{barrier}}\cdot(\frac{1}{T})
\label{Arrh}
\end{equation}
Here, $k_{CS}$ is the rate constant in ns$^{-1}$, A is the pre-exponential factor, T is the temperature in Kelvin and $E_{barrier}$ characterizes the energy barrier between the internal and surface sites. The ratio of total number of times a CS occurs to the total MD trajectory length would give the rate of exchange ($k_{CS}$) at each temperature. This allows us a way of using $E_{barrier}$ as a quantifiable `measure' of the strength between the core (internal atoms) and surface atoms i.e. an intrinsic, cluster-dependent and intensive quantity which could be correlated with the \textit{greater-than-bulk melting} character ($\nu_{GB}$). 

Shown in the inset of Fig.~\ref{Arrhenius}  are the data points and the corresponding linear least squares fit based on the \textit{`persistent internal} atom criterion \textit{(a)'}. We observe stronger fluctuations in the natural logarithm of the CS rate below the melting temperature and thus decided to perform the least squares fit to data points above the melting temperature. The obtained $E_{barrier}$ values from the fit has been tabulated in Table~\ref{ebarrier_table}. Although we do observe a clear increase in the energy barrier with increasing Al atoms in the Ga$_{(20-x)}$Al$_{x}$$^+$ cluster series, we found the \textit{`persistent internal'} atom criterion \textit{(a)} quite loose for the temperatures below the melting temperature for tagging an atom to be \textit{internal}. The high mobility of the surface atoms led to frequent expansion and shrinkage of the cluster volume as observed in the MD. Thus, although when observed over a time range as in criterion (b), the \textit{`persistent internal'} atom does extremely well to tag a particular atom as internal, it needs further refining when monitoring changes at every time step. Furthermore, from MD observations we found two ways in which the internal and surface atoms and sites were exchanged: (i) stage-wise where a surface atom became internal in subsequent MD steps, and (ii) due to parallel tempering. With increasing temperatures, we also observed changes in the number of internal and surface atoms reflecting presence of isomeric structures near the melting temperatures. We detail our observations from the MD movies in section 3 (Supplementary Information).  

\begin{figure}[t!]
\centering
\includegraphics[trim = 0cm 0cm 0cm 0cm,clip=true,width=\columnwidth]{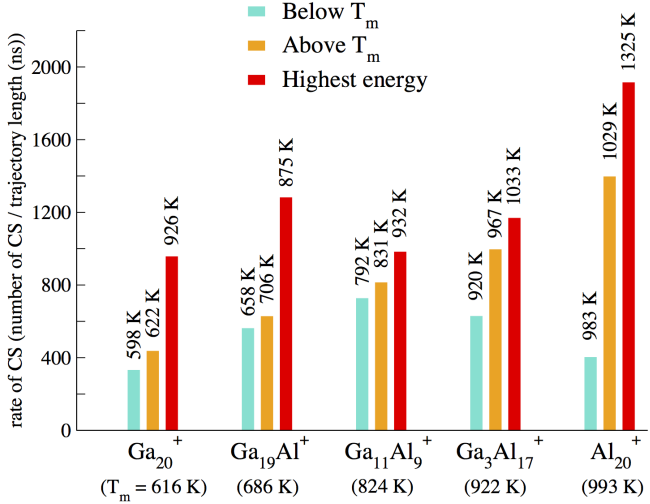}
\caption{Rate of change-swap (CS) between the internal and external atoms in the representative Ga$_{(20-x)}$Al$_{x}$$^+$ clusters. The temperatures at which the rate of CS was calculated is written above the bars.}
\label{percent}
\end{figure}
 
 \begin{figure}[t!]
\centering
\includegraphics[trim = 0cm 0cm 0cm 0cm,clip=true,width=\columnwidth]{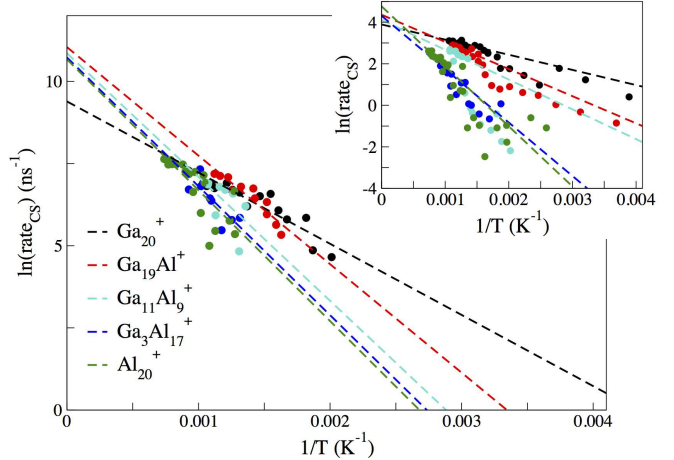}
\caption{Arrhenius plot showing the dependence of natural logarithm of the rate of CS between the internal and surface sites to the inverse of temperature. The dashed lines are the least-square straight line fits to the data points obtained from the convex hull method. The inset shows the corresponding data points obtained from the \textit{persistent internal} atom criterion (\textit{a}). The line fit has been done only for temperatures above the melting point for each cluster.}
\label{Arrhenius}
\end{figure}

To counter the problem of accurately tagging a particular atom as \textit{internal} at each MD step, we employed the convex hull technique\cite{convexhull} which for a finite set of points (in this case 20) is the union of all convex combinations of them. For each configuration of 20 points, we perform the convex hull in both two and three dimensions. In three dimensions, we disregard the atoms which form the vertices of the hull. In the two dimensions, we calculate the convex hull for the top view, front view and side view of the cluster. A cartesian coordinate (atomic site) is considered \textit{internal} if it is a non-hull vertex in all the three respective two dimensional convex hulls along with being a non-hull vertex in the corresponding three dimensional convex hull also satisfying the criterion \textit{(a)} of the \textit{`persistent internal'} atom. 

\begin{table}[b]
\centering
\caption{Comparison of the $E_{barrier}$ values obtained using the two methods described above.}
\begin{tabular}{c c c}
\hline
\hline
 & \textit{`Persistent internal'} & Convex hull\\
 & atom criterion \textit{(a)} & method\\
 \hline
Ga$_{20}^{+}$ & 729.5 & 2165.6\\
Ga\textsubscript{19}Al\textsuperscript{+} & 1306.5 & 3301.4\\
Ga\textsubscript{11}Al\textsubscript{9}\textsuperscript{+} & 1433.9 & 3767.3 \\
Ga\textsubscript{3}Al\textsubscript{17}\textsuperscript{+} & 2570.2 & 3921.3 \\
Al\textsubscript{20}\textsuperscript{+} & 2889.5 & 3982.3 \\
\hline
\end{tabular}
\label{ebarrier_table}
\end{table}

Fig~\ref{percent} shows the rate of CS for the representative clusters during the microcanonical simulations, at average  temperatures below the melting temperature, above the melting temperature and at the highest energy simulated in which case all clusters are fully \textit{liquid-like}. We have ignored those CS brought about by parallel tempering where there is a change only in the number of internal sites (isomeric) as opposed to the atom occupying the internal site as well (homotopic). In all cases, we see an increase in the rate of CS with temperature. However, we also observe that below the melting temperature the rate of CS is dependent on the amount of mixing among representative clusters. 
   
\begin{figure}[t]
\centering
\includegraphics[trim = 0cm 0cm 0cm 0cm,clip=true,width=\columnwidth]{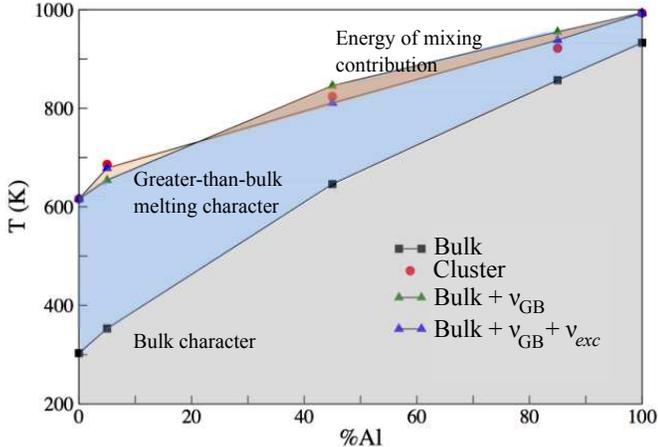}
\caption{Comparison of the DFT calculated cluster melting temperature (red) of the five representative compositions of Ga\textsubscript{(20-x)}Al\textsubscript{x}\textsuperscript{+} cluster system to that obtained considering the picture described in Eq.~\ref{model}.}
\label{model_pic}
\end{figure}

Shown in Fig.~\ref{Arrhenius} is the corresponding natural logarithm of the rate of CS calculated at average temperatures for different microcanonical trajectories. The temperatures for which CS was less than 10 discrete instances during the MD trajectory, or the contribution of PT to CS was more than 15\% have been discarded, leaving a minimum of eight data points. Shown as dashed lines are the corresponding linear least squares line fit to the values. Comparing to the inset of Fig.~\ref{Arrhenius}, we observe that the spread in data at temperatures below the melting temperature has been significantly reduced, reflecting the importance of added screening to gauge a CS situation. The slopes of the line-fit, i.e. $E_{barrier}$ for the different clusters, have been tabulated in Table~\ref{ebarrier_table}. We find that with increasing Al content in the cluster, the energy barrier for a CS situation increases. It must however be noted that the actual magnitude of $E_{barrier}$ depends not only on the number of data points and thus the fitting itself but also the method used to calculate the rate of CS. From Table~\ref{ebarrier_table}, correlating with the $\Delta T_m$ values tabulated in Table~\ref{melttemp}, we observe that there is an inverse relationship between the relative temperature difference between the cluster and the corresponding bulk Ga-Al alloy and the greater-than-bulk melting character (GB). 

Expressing $\nu_{GB} = \alpha f(E_{barrier})$, we first perform a linear least squares fit to the $E_{barrier}$ values obtained using the convex hull method for the pure phases, and follow by optimising $\alpha$. Represented as green triangles we show in Fig.~\ref{model_pic} the obtained melting temperatures of the five representative clusters when GB character has been added to the bulk melting temperatures. Furthermore, we express $\nu_{exc} = \beta E_{exc}$ and thereby correct the melting temperatures of the mixed phases which are now within 2\% error from the DFT calculated (red circles) melting temperatures, shown as blue triangles in Fig.~\ref{model_pic}. 

\section{Conclusion}
First principles calculations predict greater-than-bulk melting behaviour for the mixed gallium-aluminium bimetallic clusters: Ga\textsubscript{11}Al\textsubscript{9}\textsuperscript{+} and Ga\textsubscript{3}Al\textsubscript{17}\textsuperscript{+}. This behaviour has already been observed experimentally for Ga\textsubscript{20}\textsuperscript{+} \& Ga\textsubscript{19}Al\textsuperscript{+} and predicted for Al\textsubscript{20}\textsuperscript{+}. Moreover, the melting temperature of the Ga\textsubscript{(20-x)}Al\textsubscript{x}\textsuperscript{+} nanoalloys increases with increasing Al content. The differences between the environments of internal and surface sites of the mixed clusters has been convincingly demonstrated, a feature which seems common to the Ga\textsubscript{(20-x)}Al\textsubscript{x}\textsuperscript{+} cluster series. An Arrhenius-like analysis describes this difference as an energy barrier which gives a way to quantify the greater-than-bulk melting character of these nanoalloys. Furthermore, we propose a picture where the  Ga\textsubscript{(20-x)}Al\textsubscript{x}\textsuperscript{+} clusters are viewed as components of the corresponding bulk alloy phase with a systematic increase in the melting temperature which can be ascribed to this finite size specific energy barrier.      

\section{Acknowledgement}
The support of Royal Society of New Zealand (RSNZ) through the Marsden Fund (contract no. IRL0801) and New Zealand e-Science Infrastructure (NeSI) for computational resources is gratefully acknowledged. We would also like to thank Krista G. Steenbergen for useful discussions.  

\bibliography{ref,rsc} 
\bibliographystyle{plain}

\end{document}